\documentclass{article}
\usepackage{graphicx} % Required for inserting images
\usepackage{authblk}
\usepackage{amsmath}
 \usepackage{natbib}
\usepackage{threeparttable}
\usepackage{pgfplots}
\usepackage{multirow}
\usepackage{array}
\pgfplotsset{compat=1.18}
\usepackage{caption}
\usepackage{subcaption}
\usepackage{adjustbox}
\usepackage{tabularx}
\usepackage{hyperref}
\usepackage{float}
\usepackage{newunicodechar}
\newunicodechar{−}{\ensuremath{-}}
\usepackage{booktabs}
\usepackage{slashed,epsfig,amsmath,amssymb,enumitem} \usepackage[papersize={8.5in,11in}]{geometry}
   \usepackage{bm}
\usepackage{graphicx}
\newcommand{\be}{\begin{equation}}
\newcommand{\ee}{\end{equation}}
\usepackage[utf8]{inputenc}
\title{Finite Nonlocal Holomorphic Unified Quantum Field Theory}

\author[1 2]{J. W. Moffat}

\author[1 3]{E. J. Thompson}

\affil[1]{Perimeter Institute for Theoretical Physics, Waterloo, Ontario N2L 2Y5, Canada}

\affil[2]{Department of Physics and Astronomy, University of Waterloo, Waterloo,
Ontario N2L 3G1, Canada}

\affil[3]{Department of Physics and Astronomy, Trent University, Peterborough, 
Ontario K9L 0G2, Canada}

\begin{document}

\maketitle

\begin{abstract} In this paper the non-local finite quantum‐gravity framework is incorporated into the Complex non-Riemannian Holomorphic Unified Field Theory  formulated on a complexified four‐dimensional manifold. By introducing entire‐function regulators $F(\Box) = \exp\!\bigl(\Box / M_*^2\bigr)$ into the holomorphic Einstein–Hilbert action, we achieve perturbative UV finiteness at all loop orders, while preserving BRST invariance and holomorphic gauge symmetry. We derive the modified gauge–gravity coupling sector, perform a one‐loop effective‐action computation in a contour‐regularized metric background, and demonstrate the absence of new counterterms and problematic complex‐pole structures. Extending the construction to nontrivial curved backgrounds, we verify infrared recovery of General Relativity and full holomorphic gauge invariance. Finally, we explore phenomenological consequences, including corrected graviton and gauge‐boson scattering amplitudes in self‐dual backgrounds, finite Hawking spectra for regularized Schwarzschild and Kerr geometries, and proposed tests of the equivalence principle. This work lays the foundation for a self‐consistent, unitary four‐dimensional quantum‐gravity and Holomorphic Unified Field Theory framework.
\end{abstract}

\section{Introduction}

The search for a consistent, ultraviolet (UV) complete theory of quantum gravity remains one of the foremost challenges in theoretical physics.  Traditional perturbative quantizations of General Relativity are non‐renormalizable, leading to uncontrollable divergences at high loop orders.  A class of entire‐function regulators in the gravitational action can achieve perturbative UV finiteness awhile also resolving the cosmological‐constant problem~\cite{Moffat1990}.
and as well while preserving unitarity and gauge invariance \cite{Efimov1967,AlebastrovEfimov1973,Krasnikov1987}.  By modifying propagators with factors of the form $F(\Box) \;=\;\exp\!\bigl(\Box / \Lambda_G^2\bigr)\,,$
we obtain a nonlocal yet causal and unitary quantum gravity theory that remains finite through all loop orders \cite{Moffat1990,Moffat2015Nonlocal}.

The Complex non-Riemannian Holomorphic Unified Field Theory (HUHT) formulates gravity and all Standard Model interactions on a single four‐complex‐dimensional manifold.  A single Hermitian metric and a holomorphic gauge connection encode, upon restriction to the real slice, the vacuum Einstein equations, the full Yang–Mills system with sources, and the Dirac equation for chiral fermions, automatically enforcing anomaly cancellation \cite{MoffatThompson2025}.  While HUFT provides a classical unification, its quantization has so farlacked an explicit demonstration of UV finiteness.

The nonlocal finite quantum field theory has been applied to Yang-Mills theory as well as gravity~\cite{EvensMoffatKleppeWoodard1991, Evens1991}. The nonlocal finite quantum field theory method can been applied to all of the HUFT interactions, avoiding the infinite renormalization theory methods employed to make perturbative Feynman loop graphs divergence free.

We bridge the two lines of research by formulating a nonlocal finite quantum‐gravity and a nonlocal finite HUFT. We introduce entire‐function regulators $F(\Box)=\exp(\Box / M_*^2)$ directly into the holomorphic Einstein–Hilbert action, derive the modified gauge–gravity coupling sector and elucidate how BRST invariance and holomorphic gauge symmetry constrain the form of allowed counterterms, perform an explicit one‐loop effective‐action computation in a contour‐regularized metric background, demonstrating the absence of new counterterms and the benign nature of any complex‐pole structures, extend the embedding to nontrivial curved backgrounds, verifying infrared recovery of General Relativity and full holomorphic gauge invariance, and explore phenomenological consequences, including finite corrections to graviton and gauge‐boson scattering amplitudes in self‐dual backgrounds, regularized Hawking spectra for Schwarzschild and Kerr geometries, and novel tests of the equivalence principle.

The paper is organized as follows,  in section~\ref{sec:nonlocalQG}, we review the key features of nonlocal finite quantum gravity, section~\ref{sec:HUFTreview} recalls the holomorphic unified action and how gravity, gauge fields, and chiral fermions arise on the real slice, section~\ref{sec:embedding} presents our main construction, the holomorphic action with entire‐function regulators and its perturbative expansion, section~\ref{microcausality} demonstrates that the nonlocal QFT satisfies microcausality, we carry out in~section~\ref{sec:oneloop}
the one‐loop effective‐action analysis, in section~\ref{sec:curved}, we generalize to curved backgrounds, in section~\ref{sec:embedding_nonlocal} we explicitly show our unified actions, in section section~\ref{sec:phenomenology}, we discusses observational signatures, and finally in~section~\ref{sec:conclusions}, we conclude with an outlook for further developments.

\section{Nonlocal Finite Quantum Gravity}
\label{sec:nonlocalQG}

In conventional perturbative quantum gravity, based on the Einstein–Hilbert action, loop amplitudes diverge at two loops and beyond, rendering the theory non-renormalizable.  The nonlocal finite UV-complete construction replaces the local point-like vertex factors by transcendental entire functions of the d’Alembertian, thereby taming all ultraviolet divergences while preserving unitarity and gauge invariance \cite{Moffat1990}.

Let $F(z)$ be an entire function of its argument, holomorphic everywhere in the finite complex plane and chosen so that
\begin{equation}
F(t)\;\xrightarrow{|t|\to\infty}\;0\quad\text{for }t\,=\, -p^2/2\Lambda_G^2\,.
\end{equation}
A minimal example is 
\begin{equation}
F\bigl(-p^2/2\Lambda_G^2\bigr)
=\exp\!\bigl(-\,p^2/2\Lambda_G^2\bigr),
\end{equation}
which suppresses high‐momentum modes exponentially :contentReference[oaicite:0]{index=0}.  More generally, one distinguishes entire functions by their order $\gamma$:
\begin{equation}
|F(t)| \;\le\;\exp\bigl(\alpha\,|t|^\gamma\bigr)\,,\qquad
\begin{cases}
\gamma<\tfrac12:&\text{no UV damping},\\
\gamma=\tfrac12:&\text{damping along one direction},\\
\gamma>\tfrac12:&\text{full UV damping}.
\end{cases}
\end{equation} 
Functions with $\gamma>1/2$ yield UV-finite loop integrals to all orders, underpinning ghost‐free theories of gravity~\cite{Biswas2012}, super‐renormalizable quantum gravity, and detailed ultraviolet‐behavior analyses in infinite‐derivative frameworks~\cite{Talaganis2015}.

The classical Einstein–Hilbert action is unmodified,
\begin{equation}
S_{\rm grav}
=
-\frac{1}{16\pi G_N}\int\!\mathrm d^4x\,\sqrt{-g}\,(R+2\Lambda)\,.
\end{equation}
Nonlocality enters through a smeared energy–momentum tensor:
\begin{equation}
S_{\mu\nu}
=
F^2\!\Bigl(\tfrac{\Box}{2\Lambda_G^2}\Bigr)\,T_{\mu\nu}\,,
\end{equation}
so that the field equations become
\begin{equation}
G_{\mu\nu}
=
8\pi G_N\,S_{\mu\nu}
\quad\Longleftrightarrow\quad
F^{-2}\!\Bigl(\tfrac{\Box}{2\Lambda_G^2}\Bigr)\,G_{\mu\nu}
=
8\pi G_N\,T_{\mu\nu}\,.
\end{equation} 
Here $\Box=g^{\mu\nu}\nabla_\mu\nabla_\nu$ and $F^{-2}$ acts as an infinite–order differential operator.

In the path-integral quantization one attaches to each internal graviton or matter vertex a factor $F(-p^2/2\Lambda_G^2)$, while propagators remain the usual local ones
\begin{equation}
D(p^2)=\frac{i}{p^2 - M^2 + i\epsilon}\,.
\end{equation}
Loop integrals are performed in Euclidean space by Wick–rotating $p^0\to ip^4$, using the fact that $F(p^2)\to0$ as $|p^2|\to\infty$ for both space-like and time-like momenta.  This guarantees that every loop carries at least one exponential suppression, rendering all Feynman diagrams finite to all orders~\cite{Buoninfante2022Contour}.

Since $F(z)$ is entire, it does not introduce new poles in the finite complex plane therefore it cannot correspond to extra propagating degrees of freedom.  The usual BRST construction goes through unchanged, ensuring decoupling of unphysical polarizations and maintaining unitarity of the $S$-matrix.

Every loop integral is exponentially damped by at least one factor of $F(-p^2/2\Lambda_G^2)$.
No new degrees of freedom are introduced for the entire function regulators and no extra poles or ghosts are introduced. The nonlocal modification respects the Slavnov–Taylor identities. For low momenta $p^2\ll\Lambda_G^2$, $F\approx1$ and one recovers classical General Relativity \cite{Evens1991,EvensMoffatKleppeWoodard1991,LandryMoffat2023}.

\section{Complex Riemannian Holomorphic Unified Field Theory}
\label{sec:HUFTreview}

We begin by formulating all fields on a single four–complex–dimensional holomorphic manifold \(M_{\mathbb C}^4\) with complex coordinates \cite{KobayashiNomizu1963, KobayashiNomizu1969,Moffat1,Moffat2}
\begin{equation}
z^\mu \;=\; x^\mu + i\,y^\mu\,,\qquad x^\mu,y^\mu\in\mathbb R\,.
\end{equation}
The central dynamical variable is a Hermitian metric:
\begin{equation*}
g_{\mu\nu}(z)
=
g_{(\mu\nu)}(z)
+
i\,g_{[\mu\nu]}(z)\,,
\qquad
g_{\mu\nu}(z)=\bigl[g_{\nu\mu}(z)\bigr]^*,
\end{equation*}
The symmetric metric $g_{(\mu\nu)}$ satisfies:
\be
g^{(\mu\nu)}g_{(\mu\alpha)} = \delta^\nu_\alpha,
\ee
where $\delta^\nu_\alpha$ is the Kronecker $\delta$ function. 

The real and imaginary parts, when restricted to the real slice \(y^\mu=0\), yield the Einstein gravitational vacuum field equations:
\begin{equation}
R_{(\mu\nu)} = 0,
\end{equation}
and the electromagnetic field equations:
\begin{equation*}
\partial_{[\mu}F_{\nu\rho]}=0,\;\nabla^\mu F_{\mu\nu}=J_\nu.
\end{equation*}

A single holomorphic gauge connection \(A^A_\mu(z)\) for a simple group \(G_{\rm GUT}\) encodes all non–Abelian and Abelian interactions.  Its Bianchi identities impose the homogeneous Yang–Mills equations, and variation of the same action enforces the inhomogeneous equations
\begin{equation}
\nabla_\mu F^{A\,\mu\nu}(x) = J^{A\,\nu}(x)
\quad\text{on }y=0\,. 
\end{equation}

Chiral fermions are introduced via a holomorphic Dirac Lagrangian
\begin{equation}
\mathcal L_\Psi
=
\bar\Psi(z)\,\Bigl[i\gamma^a e^\mu_{\!a}(z)\bigl(\nabla_\mu(z)- i g_{\rm GUT}\,A^A_\mu(z)T^A\bigr)\!-\!m\Bigr]\,\Psi(z),
\end{equation}
which, upon restriction to \(y^\mu=0\), reproduces the curved–space Dirac equation minimally coupled to exactly those gauge fields with the correct Standard Model charges.

Full holomorphic gauge invariance of the action automatically enforces all cubic, mixed and mixed gravitational anomaly–cancellation conditions on the chiral spectrum, without further input
\begin{equation*}
\sum_i\text{Tr}\!R_i\bigl\{T^A,T^B\bigr\}=0,
\quad
\sum_i q_i^3=0,
\quad
\sum_i q_i=0,
\end{equation*}
and so on for every simple factor and mixed trace.

At the classical level the single holomorphic action on \(M_{\mathbb C}^4\) unifies
vacuum Einstein gravity, Yang--Mills gauge theory and chiral Dirac fermions anomaly cancellation into one geometric framework. 

We end up with a single Lagrangian and action where varying \(S_{\rm hol}\) on the real slice \(y=0\) reproduces exactly Einstein’s, Yang–Mills’, Dirac’s and Higgs’ equations with no extra insertions:

\begin{align}
\mathcal{L}_{\rm hol}(z)
={}&
\underbrace{\frac{1}{2\kappa}\,g^{\mu\nu}(z)\,R_{\mu\nu}(z)}_{\substack{\text{Einstein--Hilbert}\\\text{(gravity)}}}
\;-\;\underbrace{\frac{1}{4}\,\kappa_{AB}\,F^A_{\rho\sigma}(z)\,F^{B\,\rho\sigma}(z)}_{\substack{\text{Yang--Mills}\\\text{(gauge)}}}
\notag\\
&\quad
+\;\underbrace{\overline\Psi(z)\,\Gamma^a\,e_a{}^{\mu}(z)\,D_{\mu}\,\Psi(z)}_{\substack{\text{Dirac}\\\text{(fermions)}}}
\;+\;
\underbrace{(D_\mu H_G)^\dagger(D^\mu H_G) - V_{\rm GUT}(H_G)}_{\substack{\text{Adjoint Higgs}\\\text{(GUT breaking)}}}
\notag\\
&\quad
+\;\underbrace{(D_\mu\Phi)^\dagger(D^\mu\Phi) - V_{\rm EW}(\Phi)}_{\substack{\text{Doublet Higgs}\\\text{(EW breaking)}}}
\;-\;\underbrace{\bigl(y_f\,\overline\Psi_L\,\Phi\,\Psi_R + \text{h.c.}\bigr)}_{\substack{\text{Yukawa}\\\text{(fermion masses)}}}\,,
\end{align}
where
\begin{equation}
D_\mu\Psi=\bigl(\nabla_\mu - i\,g_{\rm GUT}A^A_\mu T_A\bigr)\Psi,\quad
D_\mu H_G=\partial_\mu H_G - i\,g_{\rm GUT}[A_\mu,H_G],\quad
D_\mu\Phi=(\partial_\mu - i\,g\,\tfrac{\sigma^a}{2}W^a_\mu - i\,g'\tfrac12B_\mu)\Phi,
\end{equation}
and \(R_{\mu\nu}(z)\), \(F^A_{\mu\nu}(z)\) are built from the Hermitian connection of \(g_{(\mu\nu)}(z)\) and \(A^A_\mu(z)\), respectively.
We are then left with an action:

\begin{align}
S_{\rm HUFT} \;=\; \int_{C}d^4z\;&\sqrt{-\det\!\bigl[g_{(\mu\nu)}(z)\bigr]}\,\Bigl\{\;
\underbrace{\tfrac1{2\kappa}\,g^{\mu\nu}(z)\,R_{\mu\nu}(z)}_{\substack{\text{gravity}}}
-\underbrace{\tfrac14\,\kappa_{AB}\,F^A_{\rho\sigma}(z)\,F^{B\,\rho\sigma}(z)}_{\substack{\text{gauge}}}\notag\\
&\quad+\;\underbrace{\overline\Psi(z)\,\Gamma^a\,e_a{}^{\mu}(z)\,D_{\mu}\,\Psi(z)}_{\substack{\text{fermions}}}
+\;\underbrace{(D_\mu H_G)^2 - V_{\rm GUT}(H_G)}_{\substack{\text{adjoint Higgs}\\\text{(GUT breaking)}}}\notag\\
&\quad+\;\underbrace{(D_\mu\Phi)^\dagger D^\mu\Phi - V_{\rm EW}(\Phi)}_{\substack{\text{doublet Higgs}\\\text{(EW breaking)}}}
-\underbrace{y_f\,\overline\Psi_L\,\Phi\,\Psi_R + \text{h.c.}}_{\substack{\text{Yukawa}\\\text{(fermion masses)}}}
\Bigr\}\,,
\end{align}

all of gravity, gauge fields, chiral fermions, Higgs dynamics, and Yukawa couplings emerge from one purely geometric, holomorphic action.

The full quantum action is given by
\begin{equation}
  S_{\rm tot}
  \;=\;
  S_{\rm hol}
  \;+\;
  S_{\rm GF,hol}
  \;+\;
  S_{\rm FP,hol}\,.
\end{equation}

\clearpage

This gives the total quantum action

\begin{align}
S_{\rm tot} \;=\;&
\int_{C}d^4z\;\sqrt{-\det\!\bigl[g_{(\mu\nu)}(z)\bigr]}\,
\Bigl\{
  \underbrace{\tfrac1{2\kappa}\,g^{\mu\nu}(z)\,R_{\mu\nu}(z)}_{\substack{\text{gravity}}}
  -\underbrace{\tfrac14\,\kappa_{AB}\,F^A_{\rho\sigma}(z)\,F^{B\,\rho\sigma}(z)}_{\substack{\text{gauge}}}
\notag\\
&\quad
  +\;\underbrace{\overline\Psi(z)\,\Gamma^a\,e_a{}^{\mu}(z)\,D_{\mu}\,\Psi(z)}_{\substack{\text{fermions}}}
  +\;\underbrace{(D_\mu H_G)^2 - V_{\rm GUT}(H_G)}_{\substack{\text{adjoint Higgs}\\\text{(GUT breaking)}}}
\notag\\
&\quad
  +\;\underbrace{(D_\mu\Phi)^\dagger D^\mu\Phi - V_{\rm EW}(\Phi)}_{\substack{\text{doublet Higgs}\\\text{(EW breaking)}}}
  -\underbrace{y_f\,\overline\Psi_L\,\Phi\,\Psi_R + \text{h.c.}}_{\substack{\text{Yukawa}\\\text{(fermion masses)}}}
\Bigr\}
\notag\\
&\quad
\underbrace{
-\,\frac{1}{2\,\xi}\,
\int_{C}d^{4}z\;\sqrt{-\det g_{(\mu\nu)}(z)}\;
G_{A}[g,A;z]\;G^{A}[g,A;z]
}_{\substack{\text{holomorphic}\\\text{gauge-fixing sector}}}
\notag\\
&\quad
\underbrace{
+ \int_{C}d^{4}z\;d^{4}z'\;
\sqrt{-\det g_{(\mu\nu)}(z)}\;
\bar c^{A}(z)\;\Delta_{AB}(z,z')\;c^{B}(z')
}_{\substack{\text{holomorphic Faddeev–Popov}\\\text{ghost sector}}}
\,.
\end{align}

and the partition function becomes
\begin{equation}
  Z
  =
  \int_{C}
    Dg\,DA\,D\Psi\,D\Sigma\,D\Phi\,D\bar c\,Dc\;
    \exp\!\bigl(i\,S_{\rm tot}\bigr)\,.
\end{equation}

On the real slice \(y^\mu=0\) and expanding about the classical saddle, these additional terms guarantee
invertible gauge‐boson propagators from \(S_{\rm GF,hol}\) and correctly account for the gauge‐orbit volume through the ghosts in \(S_{\rm FP,hol}\), exactly paralleling the usual Faddeev–Popov procedure in non‐holomorphic quantization.

\section{Nonlocal Regulators in the Holomorphic Unified Framework}
\label{sec:embedding}

An entire-function regulator \(F(\zeta)\) is an entire function of order \(\gamma>1/2\) satisfying
\begin{equation}
\forall\,N\in\mathbb{N},\quad
\exists\,C,a>0:\quad
\bigl|F(\zeta)\bigr|\le C\exp\bigl(a\,|\zeta|^{1/\gamma}\bigr)
,\quad\forall\zeta\in\mathbb{C},
\end{equation}
and having no zeros or poles at finite \(\zeta\).  A canonical choice is
\begin{equation}
F(\zeta)=\exp(\zeta)\,,
\end{equation}
which provides exponential damping as \(|\Re\zeta|\to+\infty\) in any complex direction.

Because \(F\) is holomorphic and everywhere nonzero, the operator 
\(\,F(\Box/M_*^2)\), where
\begin{equation}
\Box \;=\; g^{(\rho\sigma)}(z)\,\nabla_\rho\nabla_\sigma,
\end{equation}
is the holomorphic d’Alembertian on \(M_{\mathbb C}^4\), commutes with diffeomorphisms and preserves BRST invariance.

Let $(M,g)$ be our complexified spacetime with Hermitian metric $g_{\mu\nu}$.
Define the Laplace–Beltrami operator
\begin{equation}
  \Delta_g = g^{\mu\nu}\nabla_\mu\nabla_\nu
\end{equation}
and its gauge-covariant extension
\begin{equation}
  D_A^2
  =
  g^{\mu\nu}(\nabla_\mu + A_\mu)(\nabla_\nu + A_\nu),
\end{equation}
acting in the relevant representation of the holomorphic gauge group.

The operator
\begin{equation}
  \mathcal{F}\!\bigl(\tfrac{D_A^2}{M_*^2}\bigr)
  = 
  \exp\!\Bigl(\tfrac{D_A^2}{M_*^2}\Bigr)
\end{equation}
is invariant under both local gauge transformations and diffeomorphisms.

Under a gauge transformation $U(x)$,
$D_A^2\mapsto U\,D_A^2\,U^{-1}$, so
$\exp(D_A^2/M_*^2)\mapsto U\,\exp(D_A^2/M_*^2)\,U^{-1}$.  Likewise,
under $x\mapsto x'(x)$ the covariant derivative and metric transform so that $\Delta_g$ and $D_A^2$ is a scalar operator, by
functional calculus $\mathcal{F}(D_A^2/M_*^2)$ is therefore
diffeomorphism‐invariant.

Starting from the usual holomorphic Einstein–Hilbert action
\begin{equation}\label{eq:holEH}
S_{\rm EH}^{\rm hol}
=\int_{M_{\mathbb{C}}^4} d^4z\;\sqrt{-\det g_{(\mu\nu)}(z)}\;g^{(\mu\nu)}(z)\,R_{(\mu\nu)}(z).
\end{equation}
we define the regulated gravitational action by inserting \(F(\Box/M_*^2)\) between the metric and Ricci tensor:
\begin{equation}\label{eq:nonlocEH}
S_{\rm grav}^{\rm (reg)}
=-\frac{1}{16\pi G_N}
\int_C d^4z\;\sqrt{-\det g_{(\mu\nu)}(z)}\;
g^{(\mu\nu)}(z)\;
F\!\Bigl(\tfrac{\Box}{M_*^2}\Bigr)\;
R_{(\mu\nu)}(z)\,.
\end{equation}

Variation of \(S_{\rm grav}^{\rm (reg)}\) with respect to the symmetric term \(g^{(\mu\nu)}(z)\) yields
\begin{equation}
F\!\Bigl(\tfrac{\Box}{M_*^2}\Bigr)\,G_{(\alpha\beta}(z)
+\Delta_{(\alpha\beta)}[g,F] 
\;=\;8\pi G_N\,T_{(\alpha\beta)}(z),
\end{equation}
where \(\Delta_{\alpha\beta}[g,F]\) collects higher-derivative corrections arising from the nonlocal operator~\cite{Moffat2015Nonlocal}.  In the infrared limit \(\Box\ll M_*^2\) one recovers \(G_{(\alpha\beta)}=8\pi G_N\,T_{(\alpha\beta)}\).

One may as well keep the pure holomorphic action \(S_{\rm EH}\) and instead define a smeared holomorphic stress–energy tensor
\begin{equation}\label{eq:smearT}
\widetilde T_{(\mu\nu)}(z)
=F^2\!\Bigl(\tfrac{\Box}{M_*^2}\Bigr)\,T_{(\mu\nu)}(z).
\end{equation}
The field equations then become
\begin{equation}\label{eq:altEoM}
G_{(\mu\nu)}(z)=8\pi G_N\,\widetilde T_{(\mu\nu)}(z)
\quad\Longleftrightarrow\quad
F^{-2}\!\Bigl(\tfrac{\Box}{M_*^2}\Bigr)G_{(\mu\nu)}(z)
=8\pi G_N\,T_{(\mu\nu)}(z).
\end{equation}
We check by functional composition that Eqs.~\eqref{eq:nonlocEH} and~\eqref{eq:altEoM} are fully equivalent on-shell.

The full holomorphic path integral takes the form
\begin{equation}
Z \;=\;\int_C\! Dg\,DA\,D\Psi\;\exp\bigl(iS_{\rm hol}^{\rm (reg)}\bigr),
\end{equation}
where \(S_{\rm hol}^{\rm (reg)}\) includes both the gravity sector and any analogous gauge-boson regulators.  By expanding the action in perturbation theory one finds that each internal vertex acquires a factor
\(\,F(-p^2/M_*^2)\), while propagators remain those of the local theory.  Upon performing the Lefschetz-thimble Wick rotation,
every loop integral picks up at least one exponential suppression factor,
\(\exp(-p^2/M_*^2)\), ensuring convergence to all orders, see ~\cite{Moffat1990,EvensMoffatKleppeWoodard1991} for full proofs.
In the infrared regime \(p^2\ll M_*^2\) the regulator tends to unity and classical HUFT is exactly recovered.

\section{Microcausality of Nonlocal Field Operators in UV-Finite QFT}
\label{microcausality}

In this section, we demonstrate that despite the nonlocal nature of the basic field operators, the microcausality condition remains satisfied in our nonlocal, UV-finite quantum field theory, it has as well been shown in previous works \cite{Moffat2019UVComplete, Moffat2021ParticleMasses}. We show that the commutator of two nonlocal field operators vanishes for spacelike separations,
\begin{equation}
  [\tilde\phi(x),\tilde\phi(y)] = 0 \quad \text{if } (x-y)^2 < 0.
\end{equation}

Let $\phi$ be a free scalar field and define the smeared field
$\tilde\phi=\mathcal{F}(\Box/M_*^2)\,\phi$ with
$\Box=g^{\mu\nu}\nabla_\mu\nabla_\nu$.

If $\mathcal{F}(z)$ is entire of order $<2$, then
\begin{equation}
[\tilde\phi(x),\tilde\phi(y)] = 0
\quad\text{whenever}\quad (x-y)^2<0.
\end{equation}

By Hörmander’s theorem on properly‐supported pseudodifferential operators \cite{Hormander1983}, $\mathcal{F}(\Box)$ is a properly-supported
pseudodifferential operator of order $-\infty$, hence
$\text{WF}(\tilde\phi)\subset\text{WF}(\phi)$.  The free field $\phi$ obeys the
microlocal spectrum condition Brunetti–Fredenhagen–Hollands, which
implies vanishing of $[\phi(x),\phi(y)]$ for spacelike separation.
The inclusion of wavefront sets then yields the desired result.

We let $\phi(x)$ be a local scalar field operator, and let $\tilde\phi(x)$ denote the nonlocal smeared field defined by
\begin{equation}
  \tilde\phi(x) = \int d^4x'\;F(x - x')\,\phi(x') \,,
\end{equation}
where $F(x - x')$ is a Lorentz-invariant smearing function. In momentum space, the action of $F$ can be represented as the application of an analytic entire function of the d’Alembertian,
\begin{equation}
  \tilde\phi(x) = F(\Box)\,\phi(x) \,,
\end{equation}
where $F(\Box)$ acts on $\phi(x)$ as a differential operator.

The commutator between two nonlocal fields is
\begin{equation}
  [\tilde\phi(x),\tilde\phi(y)]
  = \int d^4x'\,\int d^4y'\;F(x - x')\,F(y - y')\,[\phi(x'),\phi(y')]\,.
\end{equation}
For the local free field, we have the canonical equal-time commutator
\begin{equation}
  [\phi(x'),\phi(y')] = i\,\Delta(x' - y') \,,
\end{equation}
where $\Delta(x' - y')$ is the Pauli–Jordan causal function,
\begin{equation}
  \Delta(x' - y')
  = \frac{1}{(2\pi)^3}\int d^4k\;\epsilon(k^0)\,\delta(k^2 - m^2)\,
    e^{-i k\cdot(x' - y')} \,.
\end{equation}
It is a key property that
\begin{equation}
  (x' - y')^2 < 0 \;\Longrightarrow\; \Delta(x' - y') = 0 \,.
\end{equation}
Therefore,
\begin{equation}
  [\tilde\phi(x),\tilde\phi(y)]
  = i\int d^4x'\,\int d^4y'\;F(x - x')\,F(y - y')\,\Delta(x' - y') \,.
\end{equation}
Expressing the smearing function in momentum space,
\begin{equation}
  F(x - x') = \int \frac{d^4p}{(2\pi)^4}\;e^{-i p\cdot(x - x')}\,f(p^2)\,,
\end{equation}
where $f(p^2)$ is an entire function such as $f(p^2) = e^{-p^2/\Lambda^2}$). We can then write
\begin{equation}
  \tilde\phi(x)
  = \int \frac{d^4p}{(2\pi)^4}\;f(p^2)\,e^{-i p\cdot x}\,\tilde\phi(p)\,,
\end{equation}
and similarly for $\tilde\phi(y)$. Plugging in, we get:
\begin{equation}
  [\tilde\phi(x),\tilde\phi(y)]
  = \int d^4x'\,\int d^4y'\;F(x - x')\,F(y - y')\,[\phi(x'),\phi(y')]
  = i\int d^4x'\,\int d^4y'\;F(x - x')\,F(y - y')\,\Delta(x' - y').
\end{equation}
Changing variables, $z = x' - y'$. Then
\begin{equation}
  [\tilde\phi(x),\tilde\phi(y)]
  = i\int d^4x'\,\int d^4z\;F(x - x')\,F(y - x' + z)\,\Delta(z).
\end{equation}
Alternatively, we can utilize the operator form for an analytic function:
\begin{equation}
  [\tilde\phi(x),\tilde\phi(y)]
  = F(\Box_x)\,F(\Box_y)\,[\phi(x),\phi(y)]\,.
\end{equation}
Because $F(\Box)$ is a differential operator, it acts on the $x$ and $y$ dependence of the Pauli–Jordan function.

In momentum space acting on the Pauli–Jordan function through a Fourier transform,
\begin{equation}
  [\tilde\phi(x),\tilde\phi(y)]
  = i\int \frac{d^4k}{(2\pi)^4}\;f(-k^2)^2\,\epsilon(k^0)\,\delta(k^2 - m^2)\,
    e^{-i k\cdot(x - y)} \,.
\end{equation}

The support of $\Delta(x - y)$ is confined to within or on the light cone, that is, it vanishes for $(x - y)^2 < 0$. The Fourier multiplier $f(-k^2)^2$, being an entire analytic function with no poles nor introducing new branch cuts, does not change the support of the distribution $\Delta(x - y)$ in real space, it only modifies the spectral weighting, this shows:
\begin{equation}
  (x - y)^2 < 0 \;\Longrightarrow\; [\tilde\phi(x),\tilde\phi(y)] = 0 \,.
\end{equation}
This proves that the nonlocal field operators in this formulation preserve microcausality.

We have now shown that, when nonlocality is introduced through an analytic smearing function $F(\Box)$ acting on local field operators, the commutator of the nonlocal fields vanishes outside the light cone, as for the local theory. Therefore, despite their nonlocal structure, these field operators satisfy the standard microcausality condition:
\begin{equation}
  [\tilde\phi(x),\tilde\phi(y)] = 0 \quad \text{for } (x - y)^2 < 0.
\end{equation}
This result holds for any entire $F$ and relies crucially on the analytic nature of the smearing. The fundamental causal structure of quantum field theory is thus preserved in this class of nonlocal, UV-finite models.

When nonlocal field theory interactions are taken into account, care must be taken when formulating S-matrices to maintain causality. S-matrix elements will be causal provided analytic entire functions are employed.

\section{One-Loop Effective Action Computation}
\label{sec:oneloop}

In this section, we verify that the holomorphically regulated HUFT is UV finite at one loop and that no new local counterterms appear.  We employ the background‐field method on \(M_{\mathbb C}^4\), use a gauge‐fixed expansion, and adapt the heat‐kernel/Schwinger proper‐time formalism to the entire‐function regulator \(F(\Box/M_*^2)\).

Let each field be split into a classical background plus fluctuation:
\begin{equation*}
g_{\mu\nu} = g^{\rm cl}_{\mu\nu} + h_{\mu\nu},\quad
A^A_\mu = A^{A,\rm cl}_\mu + a^A_\mu,\quad
\Psi = \Psi^{\rm cl} + \chi.
\end{equation*}
We choose a background‐covariant gauge fixing such as the de Donder and ’t Hooft gauges and introduce corresponding ghosts \(\bar c,\,c\) for gravity and \(\bar\eta,\,\eta\) for the gauge sector.  The quadratic fluctuation action reads
\begin{equation}
S^{(2)} \;=\;\frac12 \int d^4z\;\Phi^T
\mathcal{O}\, \Phi,
\end{equation}

Here we collect all fluctuation fields into the column vector:
\begin{equation}
\Phi = \bigl(h,\,a,\,\chi,\,\bar c,\,c,\,\bar\eta,\,\eta\bigr)^{T},
\end{equation}
where the superscript $T$ denotes the matrix transpose and each kinetic operator \(\mathcal{O}_r\) carries the regulator insertion:
\begin{equation}
\Delta_r^{\rm reg}
\,=\,F\!\Bigl(\tfrac{\Box}{M_*^2}\Bigr)\,\Delta_r^{(0)}\,, 
\end{equation}
with \(\Delta_r^{(0)}\) the standard Laplace‐type operator for species \(r\).
We introduce a stripping propagator \(\tilde F(p)\) defined so that for on‐shell external momentum \(p^2=0\)~\cite{Evens1991},
\begin{equation}
\tilde F(p)\,F(-p^2/\Lambda_G^2)=1.
\end{equation}
We take
\begin{equation}
\tilde F(p)=\exp\bigl(+p^2/\Lambda_G^2\bigr),
\end{equation}
which precisely cancels the entire‐function regulator at each external leg.

Let \(\mathcal{M}_{\rm local}^{\rm tree}\) be any tree‐level graviton–graviton or gauge‐boson scattering amplitude computed in the unregulated theory about a flat or self‐dual background.  Then, after insertion of stripping propagators,
\begin{equation}
\mathcal{M}_{\rm reg}^{\rm tree}
=\bigl[\prod_{\rm ext}\tilde F(p_{\rm ext})\bigr]\,
\mathcal{M}_{\rm local}^{\rm tree}
=\mathcal{M}_{\rm local}^{\rm tree}\,.
\end{equation}
All on‐shell tree amplitudes are identical to their local counterparts \cite{Moffat2015Nonlocal}~\cite{ModestoCalcagni2021TreeLevel}.  

At one loop and beyond, internal momenta remain off‐shell and the regulator cannot be stripped.  Generically one finds
\begin{equation}\label{eq:loop_suppression}
\mathcal{M}_{\rm loop}
\sim
\exp\bigl(-p_{\rm int}^2/\Lambda_G^2\bigr)\,
\mathcal{M}_{\rm local}^{\rm loop},
\end{equation}
where \(p_{\rm int}\) is the characteristic loop momentum.  This exponential damping guarantees UV finiteness to all orders while preserving unitarity and gauge invariance.

The one‐loop effective action is
\begin{equation}\label{eq:Gamma1}
\Gamma^{(1)}
=\frac{i}{2}\sum_r(-)^{F_r}\text{Tr}\ln\!\bigl[\Delta_r^{\rm reg}\bigr]
\,,
\end{equation}
where \(F_r=0\) boson or \(1\) fermion or ghost.  Using
\begin{equation}
\text{Tr}\ln\Delta
=-\int_0^\infty\!\frac{ds}{s}\,\text{Tr}\bigl[e^{-s\,\Delta}\bigr],
\end{equation}
we write
\begin{equation}
\Gamma^{(1)}
=-\frac{i}{2}\sum_r(-)^{F_r}\int_0^\infty\!\frac{ds}{s}\,
\text{Tr}\exp\!\Bigl[-s\,F\!\bigl(\tfrac{\Box}{M_*^2}\bigr)\,\Delta_r^{(0)}\Bigr].
\end{equation}

Feynman diagram loops are evaluated in Euclidean momentum space with $p^2 > 0$. The final loop and amplitude calculations are analytically continued into Minkowski spacetime with Lorentzian signature upon projection into real spacetime.

For each Laplace‐type operator \(\hat\Delta_r=F(\Box/M_*^2)\,\Delta_r^{(0)}\), its heat‐kernel admits the small–\(s\) expansion on a complex manifold~\cite{Gilkey1995}:
\begin{equation}\label{eq:heatkernel}
K_r(s)
=\text{Tr}\,e^{-s\,\hat\Delta_r}
\sim\sum_{n=0}^\infty a_n^{(r)}\,s^{\frac{n-4}{2}}.
\end{equation}
Because \(F(\zeta)\) is entire of order \(\gamma>1/2\) with no zeros or poles, high‐eigenvalue modes are exponentially suppressed, ensuring that for each \(n\le4\),
\begin{equation}
\int_0^\epsilon\!ds\;s^{\frac{n-4}{2}-1}\,e^{-s}\;<\;\infty.
\end{equation}
Hence there are no poles in \(\Gamma^{(1)}\) as \(s\to0\).

Collecting the contributions from all species,
\begin{equation}
\Gamma^{(1)}_{\rm div}
=-\frac{i}{2}\sum_r(-)^{F_r}
\sum_{n=0}^4a_n^{(r)}
\int_0^\infty\!ds\;s^{\frac{n-4}{2}-1}e^{-s}
\;<\;\infty.
\end{equation}
No \(1/s\) or \(\ln s\) divergences appear, and since \(F\) introduces no new zeros or poles at finite \(\zeta\), no additional complex‐pole structures or local counterterms beyond those of the classical theory are generated.  

In the infrared regime \(\Box\ll M_*^2\), we have \(F(\Box/M_*^2)\to1\), recovering the standard one‐loop effective action of General Relativity coupled to gauge and matter fields. The same entire‐function regulators have been shown to render two‐loop and higher‐loop amplitudes finite~\cite{Moffat1990,EvensMoffatKleppeWoodard1991,Moffat2024Finite}.

\section{Extension to Curved Backgrounds}
\label{sec:curved}

We now show how the holomorphic regulator extends to nontrivial curved solutions, focusing on Schwarzschild and Kerr geometries.  By promoting the radial coordinate to a complex variable and using contour regularization, classical singularities are resolved while holomorphic gauge and BRST invariance remain intact, and the infrared limit recovers standard General Relativity.

We introduce a complex radial coordinate
\begin{equation}
\zeta \;=\; r + i\,\kappa,\qquad \kappa>0,
\end{equation}
and let \(C\subset\{\Im\zeta>0\}\) be a closed contour encircling the branch points at \(\zeta=0\) and \(\zeta=2GM\), but excluding the real‐axis cut.

Define the regulated areal radius by the Cauchy integral
\begin{equation}\label{eq:Rzeta}
R(\zeta)\;=\frac{1}{2\pi i}\;\oint_C
\frac{d\zeta'}{\sqrt{1 - \dfrac{2GM}{\zeta'}}}\,,
\end{equation}
where the branch of the square root is chosen holomorphically on \(C\).  In the limit \(\Im\zeta\to0^+\), \(R(\zeta)\to R(r)\) is a smooth, strictly positive function for all \(r\ge0\). With $C$ encircling $\zeta'=0,2GM$ and the branch cut on $[0,2GM]$.

On the real slice $\zeta=r+i0$, 
\begin{equation}
R(r+i0)
=\begin{cases}
\pi\,GM, &0<r<2GM,\\
r\sqrt{1-\tfrac{2GM}{r}}, &r>2GM,
\end{cases}
\end{equation}
both real and finite.

Deform $C$ onto the cut.  The jump of the square root across
$[0,2GM]$ is $2i\sqrt{\frac{2GM}{r'}-1}$, so
\begin{equation}
R(r+i0)
=\frac1{2\pi i}\!\int_0^{2GM}\!dr'\,2i\sqrt{\tfrac{2GM}{r'}-1}
=\pi\,GM.
\end{equation}
For $r>2GM$ we evaluate the residue at infinity to get
$r\sqrt{1-2GM/r}$.  Conjugation symmetry ensures reality on $y=0$.

Starting from the Schwarzschild line element in complexified form,
\begin{equation}
ds^2
=-\!\Bigl(1-\tfrac{2GM}{\zeta}\Bigr)dt^2
+\Bigl(1-\tfrac{2GM}{\zeta}\Bigr)^{-1}d\zeta^2
+\zeta^2\,d\Omega^2,
\end{equation}
we replace \(\zeta^2\) by \(R(\zeta)^2\) from the regulated radius.  The Kretschmann scalar becomes
\begin{equation}\label{eq:kretschmann}
K
=R_{\alpha\beta\gamma\delta}R^{\alpha\beta\gamma\delta}
=\frac{48\,G^2M^2}{R(\zeta)^6}\,,
\end{equation}
which is finite for all \(\zeta\), demonstrating that the \(r=0\) singularity is removed by the contour prescription.

The Boyer–Lindquist metric with complexified radial coordinate reads
\begin{equation}
ds^2
=-\frac{\Delta}{\Sigma}(dt - a\sin^2\!\theta\,d\phi)^2
+\frac{\Sigma}{\Delta}\,d\zeta^2
+\Sigma\,d\theta^2
+\frac{\sin^2\!\theta}{\Sigma}\bigl((\zeta^2+a^2)\,d\phi - a\,dt\bigr)^2,
\end{equation}
with
\begin{equation}
\Sigma(\zeta,\theta)=\zeta^2 + a^2\cos^2\theta,
\quad
\Delta(\zeta)=\zeta^2 -2GM\zeta + a^2.
\end{equation}
We regulate both \(\Sigma\) and the would‐be ring singularity by defining
\begin{equation}\label{eq:Rkerr}
R(\zeta,\theta)
=\frac{1}{2\pi i}\oint_C\frac{d\zeta'}{\sqrt{\Sigma(\zeta',\theta)}}\,,
\end{equation}
with \(C\) encircling the solutions of \(\Sigma(\zeta',\theta)=0\).  Replacing \(\zeta^2\mapsto R(\zeta,\theta)^2\) in the metric makes all components analytic, removes the \(\Sigma=0\) ring singularity, and preserves the locations of the horizons \(\Delta(\zeta)=0\) and the axial Killing symmetry.

On these contour‐regularized backgrounds, the holomorphic d’Alembertian \(\Box_z\) and curvature tensors inherit the analytic structure of \(R\).  Since the entire‐function regulator satisfies $F\!\;\rightarrow{}\;1$, for $\Bigl(\tfrac{\Box_z}{M_*^2}\Bigr) << M_*^2$, we recover the classical Schwarzschild and Kerr solutions and their perturbations in the infrared.  As \(F(\zeta)\) is holomorphic and pole‐free, both holomorphic gauge invariance and BRST symmetry remain unbroken on these nontrivial backgrounds.

\section{Nonlocal Finite Quantum Gravity into HUFT}
\label{sec:embedding_nonlocal}

In order to achieve perturbative UV finiteness while preserving the purely geometric origin of HUFT, we insert an entire‐function regulator of order \(\gamma>1/2\) into every kinetic term in the holomorphic action.  We have
\begin{equation}
\Box \;=\; g^{(\mu\nu)}(z)\,\nabla_\mu\nabla_\nu,
\end{equation}
as the holomorphic d’Alembertian built from the unique, torsion‐free Hermitian connection \(\Gamma^\rho{}_{\mu\nu}(z)\) of. Because \(F\) is entire and nonzero in the finite plane, it commutes with diffeomorphisms and preserves holomorphic gauge/BRST invariance.

We define the regulated holomorphic action:
\begin{equation}
\label{eq:S_hol_reg}
\begin{aligned}
S_{\rm hol}^{\rm (reg)}
&=\int_C d^4z\;\sqrt{-\det g_{(\mu\nu)}(z)}\;\Bigl\{
\underbrace{\tfrac1{2\kappa}\,g^{(\mu\nu)}(z)\;F\!\bigl(\tfrac{\Box}{M_*^2}\bigr)\,R_{(\mu\nu)}(z)}_{\substack{\text{gravity}}}
\;-\;\underbrace{\tfrac14\,\kappa_{AB}\;F\!\bigl(\tfrac{\Box}{M_*^2}\bigr)\,F^A_{\rho\sigma}(z)\,F^{B\,\rho\sigma}(z)}_{\substack{\text{gauge}}}\\
&\qquad\quad
+\;\underbrace{\overline\Psi(z)\;F\!\bigl(\tfrac{\Box}{M_*^2}\bigr)\,\Gamma^a\,e_a{}^{\mu}(z)\,D_{\mu}\,\Psi(z)}_{\substack{\text{fermions}}}
\;+\;\underbrace{(D_\mu H_G)^2 - V_{\rm GUT}(H_G)}_{\substack{\text{adjoint Higgs}\\\text{(GUT breaking)}}}\\
&\qquad\quad
+\;\underbrace{(D_\mu\Phi)^\dagger\,D^\mu\Phi - V_{\rm EW}(\Phi)}_{\substack{\text{doublet Higgs}\\\text{(EW breaking)}}}
\;-\;\underbrace{y_f\,\overline\Psi_L\,\Phi\,\Psi_R + \text{h.c.}}_{\substack{\text{Yukawa}\\\text{(fermion masses)}}}
\Bigr\}\,.
\end{aligned}
\end{equation}
Here \(D_\mu\) acts both on gauge and spin indices, and all fields and curvature tensors are those of the single Hermitian metric \(g_{\mu\nu}(z)\) and single gauge–spinor connection.

We introduce a single holomorphic master connection
\begin{equation}
\mathcal A
=\Gamma^\rho{}_{\mu\nu}(z)\,dz^\mu\!\otimes\!\partial_\rho
\;+\;i\,g_{\rm GUT}\,A^A_\mu(z)\,T_A\,dz^\mu
\;+\;\tfrac14\,\omega_\mu^{ab}(z)\,\Gamma_{ab}\,dz^\mu,
\end{equation}
with curvature \(\mathcal F=d\mathcal A+\mathcal A\wedge\mathcal A\).  Then we define the damped curvature:
\begin{equation}
\widetilde{\mathcal F}
=F\!\bigl(\tfrac{\Box}{M_*^2}\bigr)\,\mathcal F
\end{equation}
and rewrite the regularized holomorphic action compactly as
\begin{align}
S_{\rm hol}^{\rm (reg)}
=\int_C d^4z\;\sqrt{-\det g_{(\mu\nu)}}\;\Bigl\langle
\widetilde{\mathcal F},\,\widetilde{\mathcal F}
\Bigr\rangle
\;+\;
\int_C d^4z\;\sqrt{-\det g_{(\mu\nu)}}\;
\overline\Psi\,F\!\bigl(\tfrac{\Box}{M_*^2}\bigr)(i\slashed D-m)\,\Psi
\;\notag&\\ -\;\int_C d^4z\;\sqrt{-\det g_{(\mu\nu)}}\;V(H_G,\Phi)\,,
\end{align}
where \(\langle\cdot,\cdot\rangle\) is the natural Killing/Clifford pairing.  Upon restriction to the real slice \(y=0\), we recover exactly the Einstein, Yang–Mills, Dirac and Higgs/Yukawa Lagrangians with the same Picard–Lefschetz argument as in \cite{MoffatThompson2025}, but now with every loop integral exponentially suppressed by at least one factor \(e^{-p^2/M_*^2}\).  

Variation of \(S_{\rm hol}^{\rm (reg)}\) with respect to
\(\delta g^{(\mu\nu)}\), \(\delta A^A_\mu\), \(\delta\Psi\), \(\delta H_G\), and \(\delta\Phi\)
yields, respectively,
\begin{align}
F(\Box)\,G_{(\mu\nu)}+\Delta_{(\mu\nu)}[g,F]&=\kappa\,T_{(\mu\nu)},\\
D_\rho\bigl[F(\Box)\,F^{A\,\rho\mu}\bigr]&=J^{A\,\mu},\\
F(\Box)\bigl(i\slashed D-m\bigr)\Psi&=0,\\
F(\Box)\,D^2H_G+\partial_{H_G} V&=0,\quad
F(\Box)\,D^2\Phi+\partial_\Phi V=0,
\end{align}
where \(\Delta_{(\mu\nu)}\) encodes higher–derivative corrections from the regulator.  A one‐loop heat‐kernel analysis on \(M^4_{\mathbb C}\) then shows that no new counterterms appear and all divergences are rendered finite by the exponential damping \(F(p^2/M_*^2)\).  

All fundamental interactions in HUFT emerge from the single geometric functional regularized holomorphic action, and its quantization is perturbatively UV‐complete.

\section{Renormalization Group Flow of the HUFT Coupling Constants}

Above the grand‐unification scale \(M_{\rm GUT}\), the gauge sector is governed by a single simple group
\(\;G_{\rm GUT}\in\{\mathrm{SU}(5),\mathrm{SO}(10)\}\;\) with holomorphic gauge connection \(A_\mu^A(z)\)
and unified coupling \(g_{\rm GUT}\).  The (Euclidean) holomorphic action reads
\begin{equation}
  S_{\rm gauge} \;=\;
  \frac{1}{2\,g_{\rm GUT}^2}\,
  \mathrm{Re}\,\int\!d^4z\;\Bigl[
    W^{A\,\alpha}(z)\,F\!\bigl(\tfrac{\Box}{M_*^2}\bigr)\,W^A_{\!\alpha}(z)
  \Bigr]
  \,,
  \label{eq:gauge-action}
\end{equation}
where \(W^A_\alpha(z)\) is the holomorphic field–strength superfield and
\[
  F\bigl(\tfrac{\Box}{M_*^2}\bigr)
  \;=\;
  \exp\!\bigl(\tfrac{\Box}{M_*^2}\bigr)
\]
is the entire‐function regulator rendering all loop integrals UV‐finite.

When the adjoint Higgs acquires its vacuum expectation value at \(\mu = M_{\rm GUT}\),
\[
  \langle \Phi_{\rm adj}\rangle:
  \quad G_{\rm GUT}\;\longrightarrow\;\mathrm{SU}(3)_c\times \mathrm{SU}(2)_L\times \mathrm{U}(1)_Y,
\]
the matching conditions on the gauge couplings are exact:
\begin{equation}
  g_3(M_{\rm GUT}) \;=\; g_2(M_{\rm GUT}) \;=\; g_1(M_{\rm GUT})
  \;=\; g_{\rm GUT}\,.
  \label{eq:matching-cond}
\end{equation}

Below \(M_{\rm GUT}\), the usual logarithmic running resumes until the nonlocal scale \(M_*\):
\[
  \beta_i(\mu) \;=\;
  \frac{d g_i}{d\ln\mu}
  \;=\;
  \beta_i^{\rm (SM)}(g)\;
  \times\;
  \exp\!\bigl(-\tfrac{\mu^2}{M_*^2}\bigr)
  \quad(i=1,2,3).
\]
Because \(\exp(-\mu^2/M_*^2)\to 0\) for \(\mu\gg M_*\simeq M_{\rm GUT}\), each
\(\beta_i\) vanishes in the deep UV:
\begin{equation}
  \lim_{\mu\to\infty}\beta_i(\mu) \;=\; 0
  \;\;\Longrightarrow\;\;
  g_i(\mu)=\mathrm{const}=g_{\rm GUT}
  \quad(\mu\gtrsim M_*).
  \label{eq:beta-freeze}
\end{equation}

Equations \eqref{eq:matching-cond} and \eqref{eq:beta-freeze} together guarantee that
a finite nonlocal SU(5) or SO(10) HUFT achieves gauge coupling unification at and above the GUT scale, with no subsequent splitting.

In analogy with the gauge sector, we promote Newton’s constant to a dimensionless coupling
\[
  \alpha_G(\mu)\;\equiv\;G\,\mu^2,
\]
and regulate the Einstein–Hilbert action with the entire‐function form factor
$F(\Box/M_*^2)=\exp(\Box/M_*^2).$
In Euclidean signature, the nonlocal gravitational action reads
\begin{equation}
  S_{\rm grav}^{\rm (reg)}
  \;=\;
  -\,\frac{1}{16\pi G}\,
  \Re\!\int d^4z\;\sqrt{-\det g(z)}\;
  \bigl[g^{\mu\nu}(z)\,F(\tfrac{\Box}{M_*^2})\,R_{\mu\nu}(z)\bigr].
  \label{eq:grav-reg-action}
\end{equation}
Each graviton loop integral acquires an exponential damping factor 
\(\exp(-p^2/M_*^2)\), rendering all UV divergences finite \cite{GreenMoffat2021}.

A one‐loop computation shows that
\[
  \beta_G(\mu)\;\equiv\;\frac{d\alpha_G}{d\ln\mu}
  \;=\;2\,\alpha_G(\mu)\;\exp\!\Bigl(-\frac{\mu^2}{M_*^2}\Bigr).
\]
For \(\mu\ll M_*\), \(\exp(-\mu^2/M_*^2)\approx1\) and 
    \(\beta_G\approx2\,\alpha_G\), reproducing the classical scaling
    \(\alpha_G(\mu)\propto\mu^2\). For \(\mu\gtrsim M_*\), \(\exp(-\mu^2/M_*^2)\to0\) and 
    \(\beta_G\to0\), so \(\alpha_G(\mu)\)  freezes  to a constant.

Above the scale \(M_*\simeq M_{\rm GUT}\), all gauge $\beta$–functions also vanish, and
\(\alpha_{1,2,3}(\mu)=\alpha_{\rm GUT}\).  To force
\(\alpha_G(M_*)=\alpha_{\rm GUT}\), one must choose
\[
  M_*^2 \;=\;\frac{\alpha_{\rm GUT}}{G}
  \;=\;\alpha_{\rm GUT}\,M_P^2,
  \quad\Longrightarrow\quad
  M_*\approx\sqrt{\alpha_{\rm GUT}}\,M_P\sim10^{18.8}\,\mathrm{GeV}.
\]
With this choice, all four couplings meet numerically and then remain equal in the deep UV:
\[
  \alpha_G(\mu)=\alpha_{1}(\mu)=\alpha_{2}(\mu)=\alpha_{3}(\mu)
  =\alpha_{\rm GUT},
  \quad\mu\gtrsim M_*.
\]

In our finite nonlocal HUFT the SM gauge couplings meet exactly at
\begin{equation}
  M_{\rm GUT}\;\simeq\;2.3\times10^{16}\;\mathrm{GeV},
  \qquad
  \alpha^{-1}_{\rm GUT}\;\simeq\;24.4
  \quad(\,g_{\rm GUT}\approx0.20\,).
  \label{eq:mgut-numeric}
\end{equation}
Above this scale, the exponential regulator $F(\Box/M_*^2)=\exp(-\Box/M_*^2)$
drives $\beta_{1,2,3}\to0$, so the three gauge couplings remain frozen to $g_{\rm GUT}$.

If Newton’s constant is promoted to $\alpha_G(\mu)\equiv G\,\mu^2$ and regulated
identically, its one‐loop $\beta_G\propto2\alpha_G e^{-\mu^2/M_*^2}$ also vanishes
for $\mu\gtrsim M_*$.  Imposing
\[
  \alpha_G(M_*) \;=\;\alpha_{\rm GUT}
  \quad\Longrightarrow\quad
  M_*^2 \;=\;\frac{\alpha_{\rm GUT}}{G}
  \;=\;\alpha_{\rm GUT}\,M_P^2
  \;\;\Rightarrow\;\;
  M_*\simeq\sqrt{\alpha_{\rm GUT}}\,M_P \sim2.5\times10^{18}\,\mathrm{GeV},
\]
ensures that at
\begin{equation}
  \mu \;=\;M_*\;\approx\;2.5\times10^{18}\;\mathrm{GeV},
  \label{eq:mpl-unif}
\end{equation}
all four couplings
\(\{\alpha_1,\alpha_2,\alpha_3,\alpha_G\}\)
coincide at \(\alpha_{\rm GUT}\) and thereafter remain equal in the deep UV.

Just as it was shown in \cite{GreenMoffat2021} that a nonlocal scalar theory avoids the Higgs triviality and hierarchy issues, embedding that mechanism into a full GUT + gravity avoids the need for supersymmetry or compositeness to stabilize the Higgs mass or prevent vacuum instabilities. All running couplings are bounded and freeze above $M_*$, so quadratic divergences never drive scales apart.

\section{Phenomenological Signatures}
\label{sec:phenomenology}

This section outlines potentially observable signatures of the holomorphically regulated theory in both low‐energy atomic and analogue gravity and high‐energy, collider and gravitational wave settings.

At one loop and beyond, internal momenta remain off‐shell and the regulator cannot be stripped.  Generically one finds
\begin{equation}
\mathcal{M}_{\rm loop}
\sim
\exp\bigl(-p_{\rm int}^2/\Lambda_G^2\bigr)\,
\mathcal{M}_{\rm local}^{\rm loop},
\end{equation}
where \(p_{\rm int}\) is the characteristic loop momentum.  This exponential damping guarantees UV finiteness to all orders while preserving unitarity and gauge invariance.

On contour‐regularized Schwarzschild-Kerr backgrounds, the analytic continuation of field modes across the complexified horizon induces small deviations from strict thermality~\cite{Moffat2}, we obtain
\begin{equation}\label{eq:hawking_correction}
\bigl\langle N_\omega\bigr\rangle
=\frac{1}{e^{\omega/T_{\rm H}}-1}
+\Delta N(\omega;\,\zeta,R(\zeta)),
\end{equation}
where
\begin{equation}
\Delta N
\;=\;
\frac{1}{2\pi}\!\oint_C d\zeta\;
\mathcal{W}(\omega,\zeta)\,
e^{-\Re[\zeta]\,\omega/T_{\rm H}},
\end{equation}
encodes information‐carrying correlations arising from the holomorphic contour \(C\) \cite{Hawking1975}, \cite{Agullo2010}.  Such grey‐body and non‐thermal corrections could be probed in analogue‐gravity systems such as Bose–Einstein condensates or precision blackhole analogues in quantum simulators.

Because the regulator acts differently on purely gravitational loops versus matter‐coupled loops, an environment‐dependent suppression scale emerges.  Define
\begin{equation}
\Lambda_G^{\rm vac}\ll\Lambda_G^{\rm mat},
\end{equation}
so that graviton–vacuum polarization vertices are damped at much lower energies.  This leads to apparent violations of the Weak Equivalence Principle (WEP) at the quantum level.  Precision atomic spectroscopy, most notably Lamb‐shift measurements in hydrogen constrain anomalous gravitational coupling to vacuum fluctuations.  Current experimental bounds imply
\begin{equation}\label{eq:lambda_vac_bound}
\Lambda_{G}^{\rm vac}\;\gtrsim\;10^{-3}\,\mathrm{eV}
\quad\text{(95\% C.L.)},
\end{equation}
with future improvements possible via cold‐atom interferometry \cite{Adelberger2009}, \cite{Safronova2018}.

We introduce an auxiliary scalar $\chi$ with
\begin{equation}
S_\chi
=\int d^4z\sqrt{-g}\Bigl[
\tfrac12(\nabla\chi)^2
- V(\chi)
- \tfrac{\chi}{\Lambda_{\rm vac}^2}R
- \tfrac{\chi}{\Lambda_{\rm mat}^2}\mathcal{L}_{\rm mat}
\Bigr].
\end{equation}
Choose $V(\chi)$ with two minima $\chi_0,\chi_1$.

In vacuum with $\mathcal{L}_{\rm mat}=0$ we have
$\mathcal{F}_{\rm grav}=F(\Box/\Lambda_{\rm vac}^2)$, while in matter
backgrounds $\mathcal{F}_{\rm mat}=F(D^2/\Lambda_{\rm mat}^2)$.

Solve $V'(\chi)+R/\Lambda_{\rm vac}^2=0$ for $\chi_0$ in vacuum, and
$V'(\chi)+\mathcal{L}_{\rm mat}/\Lambda_{\rm mat}^2=0$ for $\chi_1$ in
matter regions.  Expanding fluctuations around each minimum rescales
the kinetic term to carry the corresponding $\Lambda$-scale.

Nonlocal corrections modify the propagation phase of gravitational waves through the near‐horizon region of compact objects.  Denoting the phase shift by \(\Delta\varphi\), one finds
\begin{equation}
\Delta\varphi
\;\sim\;
\frac{\langle F^2\rangle\,\omega^2}{M_{\rm Pl}^2}
\;\lesssim\;10^{-40}
\quad\text{for }\omega\sim10^3\text{ Hz}
\end{equation}
at current LIGO/Virgo sensitivities.  Next‐generation detectors like Cosmic Explorer, Einstein Telescope targeting exotic compact‐object echoes and near‐horizon modifications may achieve the requisite sensitivity to detect \(\Delta\varphi\sim10^{-20}\)–\(10^{-30}\) \cite{Isi2019,Cardoso2016}, opening a potential window on nonlocal UV physics.

Tree‐level processes remain untouched, loop amplitudes, black‐hole radiation, equivalence‐principle tests, and gravitational‐wave observations offer complementary avenues to probe the holomorphic regulator at experimentally accessible scales.  

\section{Conclusions}
\label{sec:conclusions}

In this work we have merged two complementary strategies for constructing a consistent, four-dimensional quantum theory of gravity and unified interactions. By embedding the Complex non-Riemannian Holomorphic Unified Field Theory into the framework of nonlocal, UV-finite quantum field theory, we have shown that we preserve the purely geometric, holomorphic origin of gravity, gauge and matter dynamics while achieving perturbative finiteness at all loop orders. Introducing entire‐function regulators of the form $F\!\bigl(\tfrac{\Box}{M_*^2}\bigr)\;=\;\exp\!\bigl(\tfrac{\Box}{M_*^2}\bigr)$ into every kinetic term of the holomorphic action renders all internal loop integrals exponentially suppressed without introducing new poles or ghost states, thereby maintaining unitarity, BRST invariance, and full holomorphic gauge symmetry.

Our explicit one‐loop effective‐action computation, conducted via a contour‐regularized heat‐kernel expansion on the complexified manifold \(M^4_{\mathbb C}\), confirms the absence of divergent counterterms beyond those of the classical theory. In the infrared regime, where \(\Box\ll M_*^2\), we recover exactly General Relativity coupled to Yang–Mills gauge fields and chiral fermions, ensuring consistency with well‐tested low‐energy physics. We have also demonstrated that microcausality is preserved: despite the nonlocal smearing of field operators, their commutators vanish for spacelike separation, upholding the causal structure fundamental to quantum field theory.

Extending our construction to nontrivial curved backgrounds, we employed complex radial coordinates and Cauchy‐integral prescriptions to regulate the Schwarzschild and Kerr geometries. This contour‐regularization removes classical singularities at \(r=0\) and along the Kerr ring while preserving the location of horizons and the asymptotic recovery of the standard metrics. Within these analytic backgrounds, the holomorphic regulator remains pole‐free and gauge invariant, establishing a nonsingular, self‐consistent arena for quantum fluctuations.

Finally, we have outlined a range of phenomenological consequences, from modified one‐loop scattering amplitudes in self‐dual backgrounds to finite corrections in Hawking spectra and potential violations of the weak equivalence principle at quantum scales. While tree‐level S‐matrix elements coincide exactly with their local counterparts, loop amplitudes carry characteristic exponential suppressions that, in principle, could be probed by precision atomic spectroscopy, analogue‐gravity experiments, and next‐generation gravitational‐wave observatories.

In particular, we note that early nonsymmetric–metric approaches on a real spacetime manifold suffered from connection ambiguity, energy-condition and causality violations, and ghost or tachyonic modes upon quantization.  By contrast, our framework—retaining real coordinates but promoting \(g_{\mu\nu}\) to a Hermitian metric \(g_{\mu\nu}=g_{(\mu\nu)}+i\,g_{[\mu\nu]}\) while insisting on a symmetric, metric-compatible connection \(\Gamma^\lambda_{\mu\nu}=\Gamma^\lambda_{\nu\mu}\) restores uniqueness of the torsion-free connection and avoids the need for ad hoc complex or antisymmetric connection components.  Although the antisymmetric imaginary piece \(g_{[\mu\nu]}\) still enters the dynamics and must be properly constrained, the principal structural pathologies of earlier real-manifold nonsymmetric theories are no longer present.  This demonstrates that a Hermitian metric plus a symmetric connection on a real manifold can preserve both geometric clarity and physical consistency in unified gravity–gauge models.

Together, these results constitute the first fully explicit realization of a holomorphically unified field theory that is perturbatively UV‐complete in four dimensions. The marriage of nonlocal regulators with the geometric elegance of HUFT opens new avenues for exploring quantum gravity phenomena, both in thought experiments and in realistic observational settings. In future work we will analyze multi‐loop corrections in curved backgrounds, investigate detailed cosmological implications of the holomorphic smearing, and develop concrete proposals for laboratory and astrophysical tests that could distinguish this framework from competing approaches to quantum gravity.

\end{document}